\documentclass[aps,pra,twocolumn,superscriptaddress,showpacs,10pt]{revtex4-1}

\usepackage{amsmath}
\usepackage{amssymb}
\usepackage{graphicx}
\usepackage{color}
\usepackage[implicit=true,colorlinks=true,linkcolor=blue,
citecolor=blue,urlcolor=blue]{hyperref} 

\newcommand{\dd}{\mathrm{d}}
\newcommand{\ii}{\mathrm{i}}
\newcommand{\ee}{\mathrm{e}}
\newcommand{\PartDeriv}[2]{\frac{\partial #1}{\partial #2}}

\newcommand{\ket}[1]{|#1\rangle}

\newcommand{\bomega}{{\Omega}}
\newcommand{\wigcal}{\mathcal{W}_\varepsilon}

\begin{document}

\title{Tunneling through a parabolic barrier viewed from Wigner phase space}

\author{D.\,M.~Heim}
\author{W.\,P.~Schleich}
\affiliation{Institut f{\"u}r Quantenphysik and Center for Integrated Quantum Science and Technology (IQ$^{ST}$), Universit{\"a}t Ulm, D-89069 Ulm, Germany}

\author{P.\,M.~Alsing}
\affiliation{Information Directorate, Air Force Research Laboratory, Rome, NY 13441, USA}

\author{J.\,P.~Dahl}
\affiliation{Chemical Physics, Department of Chemistry, Technical University of Denmark, DTU 207, DK-2800 Kgs. Lyngby, Denmark}

\author{S.~Varro}
\affiliation{Wigner Research Centre for Physics, Hungarian Academy of Sciences, Institute for Solid State Physics and Optics, 1525 Budapest, Hungary}

\date{\today}

\begin{abstract}
We analyze the tunneling of a particle through a repulsive potential resulting from an inverted harmonic oscillator in the quantum mechanical phase space described by the Wigner function. In particular, we solve the partial differential equations in phase space determining the Wigner function of an energy eigenstate of the inverted oscillator. The reflection or transmission coefficients $R$ or $T$ are then given by the total weight of all classical phase space trajectories corresponding to energies below, or above the top of the barrier given by the Wigner function.
\end{abstract}

\pacs{03.65.-w, 03.65.Xp, 03.65.Nk, 03.65.Ca}

\maketitle

\begin{figure}[!ht]
 \includegraphics{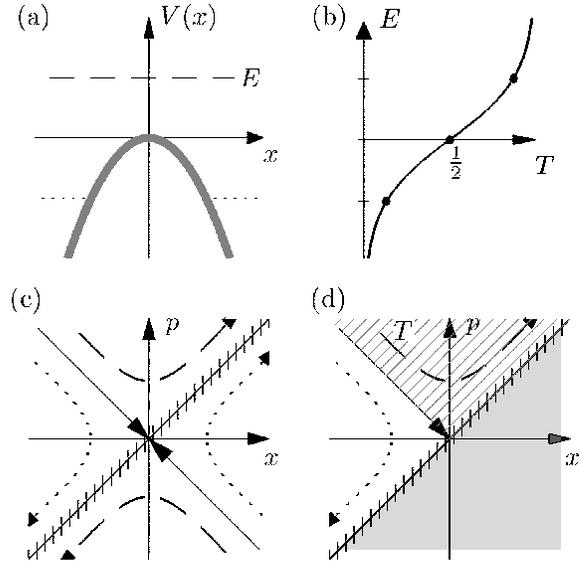}
 \caption{Tunneling coefficient $T$ of an energy eigenstate of eigenvalue $E$ through a parabolic barrier (a) in its dependence on $E$ (b) explained in terms of classical phase-space trajectories (c) subjected to the boundary conditions of a particle coming from the left (d). For three different energies -- below, at the top of, and above the barrier -- we depict the classical phase-space trajectories (c) which are either being reflected from, stopping at the top of, or going above the potential hill, respectively. The crossed line represents the separatrix in phase space separating the trajectories coming from the left and from the right. Hence, under normal scattering situations only half of phase space is accessible depicted in (d) for a particle approaching from the left. The quantum mechanical transmission curve (b) is due to the quantum mechanical weight of all classical trajectories going above the barrier provided by the Wigner function.}
 \label{fig:figure1}
\end{figure}

Tunneling \footnote{For the different aspects of tunneling see for example P. Hanggi, Z. Phys. B \textbf{68}, 181 (1987); W. H. Miller and T. F. George, J. Chem. Phys. \textbf{56}, 5668 (1972);  M. Kleber, Phys. Rep. {\bf 236}, 331 (1994); J. Ankerhold, F. Grossmann and D. Tannor, Chem. Phys. \textbf{1}, 1333 (1999); and M. Razavy, \emph{Quantum Theory of Tunneling} (World Scientific, Singapore, 2003).} of a particle through a barrier is one of the striking phenomena of quantum mechanics~\cite{Bohm1989}. In the special case of a repulsive quadratic potential, corresponding for example to an inverted harmonic oscillator~\cite{note_barton_brout} shown in Fig.~\ref{fig:figure1}(a), the transmission coefficient $T$ takes the form~\cite{kemble:1935}
\begin{equation}
 T=\frac{1}{1+\ee^{-2\pi \varepsilon}}
 , \label{eq:T:beginning}
\end{equation}
depicted in Fig.~\ref{fig:figure1}(b). Here $\varepsilon \equiv E/(\hbar \bomega)$ is the scaled energy which is the ratio of the eigenvalue $E$ and the natural energy parameter $\hbar \bomega$, where $\bomega$ is the steepness of the quadratic barrier and $\hbar$ denotes the Planck constant divided by $2\pi$.

The expression Eq.~\eqref{eq:T:beginning} has played a crucial role in the context of nuclear fission~\cite{hill:1953}. It usually emerges~\cite{hill:1953} from a semiclassical analysis~\cite{bender:1999,ford:1959} of the Schr\"odinger equation of the inverted harmonic oscillator~\cite{note_barton_brout}. However, in the present Brief Report we rederive Eq.~\eqref{eq:T:beginning} from quantum phase space using the Wigner distribution function~\cite{see:schleich}. In particular, we show that Eq.~\eqref{eq:T:beginning} corresponds to the quantum mechanical weight of all \emph{classical} trajectories~\footnote{The role of classical trajectories in the description of tunneling using the semi-classical propagator has been topic of many publications, see for example D. W. McLaughlin, J. Math. Phys. \textbf{13}, 1099 (1972); S Keshavamurthy and W. H. Miller, Chem. Phys. Lett \textbf{218}, 189 (1994); F. Grossmann and E. Heller, Chem. Phys Lett. \textbf{241}, 45 (1995); J. Ankerhold and H. Grabert, Europhys Lett. \
textbf{47}, 285 (1999); and F. Grossmann, Phys. Rev. Lett. \textbf{85}, 903 (2000).} that have sufficient energy to go above the barrier.

This result is counterintuitive since in the standard formulation~\cite{Bohm1989} of quantum mechanics \`a la Heisenberg and Schr\"odinger an energy eigenstate does not contain energies other than the eigenvalue. In contrast, the Wigner function~\cite{see:schleich} of such a state relies on the trajectories of all energies, however with positive or negative weights.

We emphasize that the Wigner function of tunneling in the inverted harmonic oscillator has also been analyzed in Ref.~\cite{balazs:1990}. The authors of this paper first derive the quadrature representation of the energy eigenfunctions and then perform the integral in the definition of the Wigner function. In contrast, we start from the two partial differential equations \cite{dahl:1983,hug:1998} determining the Wigner function from phase space. Therefore, we find the Wigner function without ever going through the wave function. This approach is not only direct but also yields immediately the proposed interpretation of the tunneling coefficient.

We study the tunneling of a particle of mass $M$ through a quadratic barrier of steepness $\bomega$ expressed by the Hamiltonian
\begin{equation}
 {H} \equiv \frac{{p}^2}{2M}-\frac{1}{2}M\bomega^2{x}^2
 . \label{eq:hamilton:prime:real}
\end{equation}
Here $x$ and $p$ denote the position and the coordinate of the particle.

For this purpose we consider the Wigner function~\cite{see:schleich}
\begin{equation}
  W(x,p)\equiv\frac{1}{2\pi\hbar}\int_{-\infty}^{\infty}\dd y\ \ee^{-\ii\, py/\hbar}\ \psi^*\left(x-\frac{y}{2}\right)\psi\left(x+\frac{y}{2}\right)
  \label{eq:wigner:def}
\end{equation}
of an energy eigenstate $\ket{E}$ of $\hat{H}$ with wave function $\psi_E=\psi_E(x)$. However, instead of solving first the time independent Schr\"odinger equation $\hat{H} \psi_E = E \psi_E$ for $\psi_E$ and then performing the integration in Eq.~\eqref{eq:wigner:def} pursued in Ref.~\cite{balazs:1990}, we analyze the partial differential equations~\cite{dahl:1983,hug:1998}
\begin{equation}
 \left[\frac{p}{M}\frac{\partial}{\partial x}+M\bomega^2 x\frac{\partial}{\partial p}\right]W_{E}(x,p)=0
 \label{eq:dgl:liouville}
\end{equation}
and
\begin{eqnarray}
 \left\{\left[\frac{p^2}{2M}-\frac{1}{2}M \bomega^2x^2\right]-\frac{\hbar^2}{8}\left[\frac{1}{M}\frac{\partial^2}{\partial x^2}-M \bomega^2 \frac{\partial^2}{\partial p^2}\right]\right\}\nonumber \\ [1ex]
 \times W_{E}(x,p)=E W_{E}(x,p)\hspace{2cm}\label{eq:dgl:wigner:orig}
\end{eqnarray}
for the Wigner function in phase space. We emphasize that Eqs.~\eqref{eq:dgl:liouville} and \eqref{eq:dgl:wigner:orig} are exact for the inverted harmonic oscillator.

The classical Liouville equation~\eqref{eq:dgl:liouville} implies that $W_E$ is constant along the classical phase space trajectories of a fixed energy $H$ given by Eq.~\eqref{eq:hamilton:prime:real} and shown in Fig.~\ref{fig:figure1}(c), that is
\begin{equation}
 W_E(x,p)=\mathcal{W}_{{E}/{(\hbar \bomega)}}\left( \frac{H(x,p)}{\hbar \bomega} \right)
 . \label{eq:wigner:scaled}
\end{equation}

Next we take into account the boundary conditions associated with a scattering process. Two distinct possibilities offer themselves: (i) the particle approaches the barrier from the left, or (ii) it impinges from the right.

The two cases manifest themselves in different classical phase space trajectories. Whereas the situation (i) is described by the trajectories in the domain above the separatrix
\begin{equation}
 p=M \bomega x
 , \label{eq:separatrix}
 \end{equation}
depicted in Fig.~\ref{fig:figure1}(d), the case (ii) covers the area below it.

Hence, for a particle coming from the left, the Wigner function $W_E^{(l)}$ of an energy eigenstate reads
\begin{subequations}
\begin{equation}
 {W}^{(l)}_E (x,p) = \mathcal{W}_{{E}/{(\hbar \bomega)}}\left( \frac{H(x,p)}{\hbar \bomega} \right) \Theta(p-M \bomega x)
 , \label{eq:wigner:l}
\end{equation}
where $\Theta$ denotes the Heaviside step function. Hence, only the classical trajectories above the separatrix contribute to the Wigner function as shown in Fig.~\ref{fig:figure1}(d).

Likewise, for a particle approaching from the right we find
\begin{equation}
 {W}^{(r)}_E (x,p) = \mathcal{W}_{{E}/{(\hbar \bomega)}}\left( \frac{H(x,p)}{\hbar \bomega} \right) \Theta(M \bomega x - p)
 . \label{eq:wigner:r}
\end{equation}
\label{eq:wigner:ansatz}
\end{subequations}
With the help of the familiar identity 
\begin{equation}
  x\delta(x)=0
  \label{eq:delta} 
\end{equation}
for the Dirac delta function it is easy to verify that both expressions satisfy the Liouville equation \eqref{eq:dgl:liouville} as long as the function $\wigcal$ is differentiable. The form of $\mathcal{W}_\varepsilon=\mathcal{W}_\varepsilon(\eta)$ corresponding to the scaled eigenvalue $\varepsilon \equiv E/(\hbar \Omega)$ in its dependence on the dimensional energy
\begin{equation}
 \eta \equiv \frac{H(x,p)}{\hbar \bomega} \equiv \frac{1}{\hbar \bomega} \left[ \frac{p^2}{2M} - \frac12 M \bomega^2 x^2 \right]
\end{equation}
of a classical trajectory is then determined by the Schr\"odinger equation \eqref{eq:dgl:wigner:orig} in phase space. Indeed, when we substitute the ansatz Eq.~\eqref{eq:wigner:ansatz} into Eq.~\eqref{eq:dgl:wigner:orig} we arrive at the ordinary differential equation
\begin{equation}
 \eta \frac{\dd^2 \wigcal}{\dd \eta^2} + \frac{\dd \wigcal}{\dd \eta} - 4 (\varepsilon - \eta) \wigcal = 0
 . \label{eq:dgl:wigner:scaled}
\end{equation}
Again we have made use of Eq.~\eqref{eq:delta}. It is remarkable that Eq.~\eqref{eq:dgl:wigner:scaled} is independent of the Heaviside step function. 

In order to solve Eq.~\eqref{eq:dgl:wigner:scaled} we make a Fourier ansatz
\begin{equation}
   \wigcal(\eta) \equiv \int_{\tau_1}^{\tau_2} \dd \tau\, w_\varepsilon(\tau)\, \ee^{\ii \eta \tau}
  \label{eq:wigner:fourier:ansatz}
\end{equation}
where the limits $\tau_1$ and $\tau_2$ of the integration will be determined in a way as to simplify the differential equation for $w_\varepsilon=w_\varepsilon(\tau)$ resulting from \eqref{eq:dgl:wigner:scaled}. Indeed, with the integral relation
\begin{equation}
 \eta \wigcal(\eta) = \int_{\tau_1}^{\tau_2} \dd \tau\, w_\varepsilon (\tau)\frac{1}{\ii} \PartDeriv{}{\tau} \left( \ee^{\ii \eta \tau} \right)
\end{equation}
and integration by parts we establish the identity
\begin{eqnarray}
 \eta \wigcal(\eta) &=& \frac{1}{\ii} \left[ w_\varepsilon (\tau_2)\, \ee^{\ii \eta \tau_2} -
  w_\varepsilon (\tau_1)\, \ee^{\ii \eta \tau_1}\right] \nonumber \\
  && - \frac{1}{\ii}\int_{\tau_1}^{\tau_2} \dd \tau\, \frac{\dd }{\dd \tau}\left[w_\varepsilon(\tau) \right] \ee^{\ii \eta \tau}
  . \label{eq:etaW}
\end{eqnarray}

Similarly, we obtain
\begin{eqnarray}
 \eta \frac{\dd^2 \wigcal}{\dd \eta^2} &=& - \frac{1}{\ii} \left[ w_\varepsilon (\tau_2) {\tau_2}^2 \, \ee^{\ii \eta \tau_2} -
  w_\varepsilon (\tau_1) {\tau_1}^2 \, \ee^{\ii \eta \tau_1}\right] \nonumber \\
  && + \frac{1}{\ii}\int_{\tau_1}^{\tau_2} \dd \tau\, \frac{\dd }{\dd \tau} \left[ \tau^2 w_\varepsilon(\tau) \right] \ee^{\ii \eta \tau}
  \label{eq:etaWdd}
\end{eqnarray}
and
\begin{equation}
 \frac{\dd \wigcal}{\dd \eta} = \int_{\tau_1}^{\tau_2} \dd \tau\, \ii \tau\,w_\varepsilon(\tau)\, \ee^{\ii \eta \tau}
 ,
\end{equation}
which when substituted into \eqref{eq:dgl:wigner:scaled} yields the ordinary differential equation
\begin{equation}
 \ii (4-\tau^2) \frac{\dd w_\varepsilon}{\dd \tau} - (\ii \tau + 4 \varepsilon) w_\varepsilon = 0
 \label{eq:dgl:ordinary}
\end{equation}
of first order for $w_\varepsilon = w_\varepsilon(\tau)$. Here we have made the choice $\tau_1 \equiv -2$ and $\tau_2 \equiv 2$ which eliminates the boundary terms in Eqs.~\eqref{eq:etaW} and \eqref{eq:etaWdd}.

By direct differentiation we can verify that 
\begin{equation}
 w_\varepsilon (\tau) \equiv A (4-\tau^2)^{-1/2} \exp \left[ -\ii \varepsilon \ln \left(\frac{2+\tau}{2-\tau} \right) \right]
 \label{eq:w_epsilon}
\end{equation}
is a solution of Eq.~\eqref{eq:dgl:ordinary} and the constant
\begin{equation}
 A \equiv \frac{1}{\pi}
\end{equation}
of integration follows from the normalization condition
\begin{equation}
 1 = \int_{-\infty}^{+\infty} \dd \eta \, \wigcal
 , \label{eq:wigner:norm}
\end{equation}
which is a consequence of the Fourier ansatz Eq.~\eqref{eq:wigner:fourier:ansatz}.

As a result, the crucial part $\wigcal$ of the Wigner function of an energy eigenstate of the inverted harmonic oscillator reads
\begin{equation}
 \wigcal(\eta)=\frac{1}{\pi} \int_{-2}^{+2} \dd \tau \frac{\exp\left[-\ii \varepsilon \ln \left(\frac{2+\tau}{2-\tau} \right)\right]}{\sqrt{4-\tau^2}}\, \ee^{\ii \eta \tau}
 . \label{eq:wigner:fourier}
\end{equation}
Since our ultimate goal is to calculate the transmission and reflection coefficients we do not discuss this expression in more detail~\footnote{With the transformation $\tau \equiv 4t-2$ this integral reduces to the integral representation of the Kummer function.} but emphasize that $\wigcal$ is real and satisfies the symmetry relation
\begin{equation}
 \mathcal{W}_{-\varepsilon}(-\eta)=\wigcal(\eta)
 . \label{eq:symmetry}
\end{equation}

\begin{figure}[!hbt]
 \includegraphics{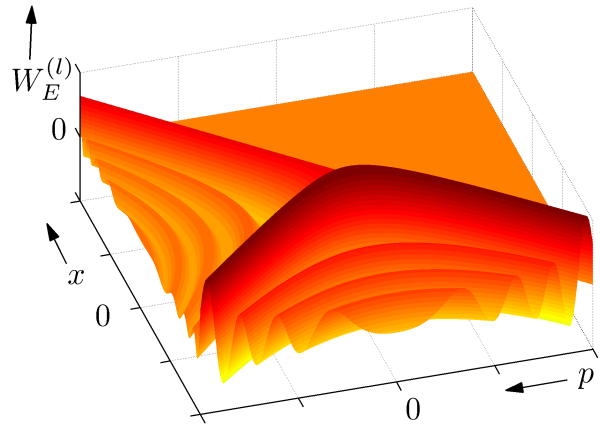}
 \caption{The Wigner function $W^{(l)}_E$ of an energy eigenstate of the inverted harmonic oscillator for the dimensionless energy $\varepsilon=-0.4$ that is below the top of the barrier and corresponding to a particle coming from the left. Due to this boundary condition the Wigner function vanishes in the phase-space domain above the separatrix (right upper part). It is constant along the classical trajectories with a dominant maximum at the trajectory corresponding to the energy eigenvalue (front). However, the Wigner function also reaches into the domain of trajectories traversing the barrier (left). The total weight underneath the Wigner function in this realm represents the tunneling coefficient.}
 \label{fig:figure2}
\end{figure}

In Fig.~\ref{fig:figure2} we depict the Wigner function of the energy eigenstate of an inverted harmonic oscillator below the top of the barrier subjected to the initial condition of the particle approaching from the left represented by Fig.~\ref{fig:figure1}(d). In order to bring out the characteristic features we have rotated the phase space by 90$^{\circ}$. The wave fronts in the foreground correspond to the phase-space trajectories of particles coming in and being reflected from the barrier. The waves on the upper left of the figure with much reduced amplitude represent particles going above the barrier. In the flat part above the separatrix extending to the right upper corner the Wigner function vanishes as to obey the boundary condition corresponding to the gray shaded area in Fig.~\ref{fig:figure1}(d).

We now turn to the evaluation of the transmission probability $T$ which is given by the integral
\begin{equation}
 T \equiv \int_0^\infty \dd \eta \, \wigcal (\eta)
 \label{eq:T:def}
\end{equation}
of the Wigner function over positive values of $\eta$ only.

With the help of the identity
\begin{equation}
 \int_0^\infty \dd \eta \, \ee^{\ii \eta \tau} = \pi \delta(\tau)+\ii\, {\color{black}\mathcal{P}\left( \color{black}\frac{1}{\tau} \color{black}\right)}
\end{equation}
we find from Eq.~\eqref{eq:wigner:fourier} the expression
\begin{equation}
 T = \frac12 + \frac{\ii}{4 \pi} \mathcal{P} \left[ \int_{-\infty}^{+\infty} \dd \xi \, \frac{\exp(-\ii \varepsilon \xi)}{\sinh(\xi/2)} \right]
 ,
\end{equation}
where we have introduced the new integration variable $\xi \equiv \ln \left[(2+\tau)/(2-\tau) \right]$ and $\mathcal{P}$ denotes the Cauchy principal part.

Due to the antisymmetry of $\sinh(\xi/2)$ only the imaginary part survives, that is
\begin{equation}
 T = \frac{1}{2} \left[1 + \frac{1}{\pi} \int_0^\infty \dd \xi\, \frac{\sin(\varepsilon \xi)}{\sinh(\xi/2)} \right]
\end{equation}
and with the integral relation~\cite{gradshteyn:1980}
\begin{equation}
 \int_0^\infty \dd \xi \, \frac{\sin(\varepsilon\, \xi)}{\sinh ( \beta\, \xi)} = \frac{\pi}{2\beta} \tanh \left(\frac{\varepsilon\pi}{2\beta}	\right)
\end{equation}
we finally arrive at
\begin{equation}
 T = \frac{1}{2} \left[ 1+\tanh(\varepsilon \pi) \right]
 ,
\end{equation}
in complete agreement with Eq.~\eqref{eq:T:beginning}.

We conclude by noting that the reflection coefficient $R$ follows from the identity
\begin{equation}
 R+T=1,
\end{equation}
together with the normalization condition Eq.~\eqref{eq:wigner:norm} and the definition Eq.~\eqref{eq:T:def} of $T$, as
\begin{equation}
 R\equiv \int_{-\infty}^{0} \dd \eta\, \wigcal(\eta).
\end{equation}

Therefore, it is the total quantum mechanical weight of the classical trajectories that are reflected, that is all energies below the maximum of the barrier.

\begin{figure}[t]
 \includegraphics{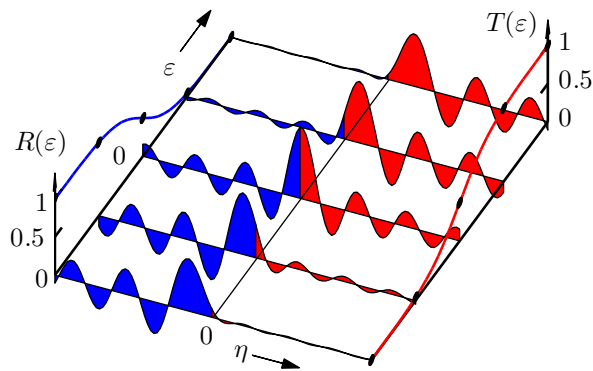}
 \caption{Reflection and transmission coefficients $R=R(\varepsilon)$ (left) or $T=T(\varepsilon)$ (right) in their dependence on the energy eigenvalue $\varepsilon$ originating from the total Wigner function weight of all classical trajectories with energies $\eta$ below or above the top of the barrier, respectively. The symmetry relation Eq.~\eqref{eq:symmetry} giving rise to $R=1-T$ is apparent.}
 \label{fig:figure3}
\end{figure}

In summary, we have rederived the familiar reflection and the tunneling coefficients $R$ and $T$ for an inverted harmonic oscillator using the corresponding Wigner function. This approach shows that $R$ and $T$ represent the quantum mechanical weights given by the Wigner function of all classical trajectories that are reflected from or transverse the barrier as indicated in Fig.~\ref{fig:figure3}. The weight of a given trajectory follows from an ordinary differential equation of second order which we have solved using a Fourier representation of the Wigner function.

Here we have not analyzed in detail the structure of the resulting differential equation of first order given by Eq.~\eqref{eq:dgl:ordinary} in the complex plane. It suffices to say, that the origin of Eq.~\eqref{eq:T:beginning} is deeply rooted in the two poles at $\tau=-2$ and $\tau=+2$ giving rise to a logarithmic phase singularity~\cite{berry:1992} contained according to Eq.~\eqref{eq:w_epsilon} in the kernel $w_\varepsilon$ of the Fourier representation of $\wigcal$.

Such singularities also appear~\cite{note_barton_brout,balazs:1990} in the quadrature representation of the energy wave function and manifest themselves in the Unruh effect~\cite{unruh:1976}, the Hawking radiation~\cite{kiefer:2009} or optical analogues~\cite{leonhardt:2002:laboratory} of event horizons of black holes. To identify in the complex plane the crucial contributions to the integral Eq.~\eqref{eq:wigner:fourier:ansatz}, or to elaborate on the importance of the logarithmic singularity, and to compare and contrast the similarities and differences between tunneling and particle creation goes beyond the scope of this Brief Report and will be the topic of a future publication.

\begin{acknowledgments}
We thank J. Ankerhold, C. Bender, I. Bialynicki-Birula, F. Grossmann, P. H\"anggi, C. Lasser, U. Leonhardt, M. Shapiro, and U. Wei\ss\ for many fruitful discussions. Moreover, we appreciate the financial support by the DFG in the framework of the SFB/TRR-21. PMA would like to acknowledge the support of the Air Force Office of Scientific Research (AFOSR) for this work. Any opinions, findings and conclusions or recommendations expressed in this
material are those of the author(s) and do not necessarily reflect the views of AFRL. SV thanks the Deutscher Akademische Austausch Dienst (DAAD) for the Research Fellowship [No. A/12/01761]. The partial support  by the Hungarian National Scientific Research Foundation OTKA, Grant No. K 104260, and the National Development Agency, Grant No. ELI\_ 09-1-2010-0010, Helios Project is also acknowledged.
\end{acknowledgments}


%
\end{document}